\documentclass[a4paper, fleqn, usenatbib]{mnras}
\pdfoutput=1
\usepackage{mathptmx}

\usepackage{graphicx}	
\usepackage{amsmath}	
\usepackage{amssymb}	
\usepackage{color}
\usepackage{enumitem}

\title[Clustering in the Cosmic Web]{Cosmic Web Type Dependence of Halo Clustering}

\author[J. D. Fisher, A. Faltenbacher]{J. D. Fisher\thanks{E-mail: justindavidfisher@gmail.com}, A. Faltenbacher\\
Department of Physics, University of Witwatersrand, Braamfontein, 2000, South Africa}

\date{Accepted XXX. Received YYY; in original form ZZZ}

\pubyear{2015}

\begin{document}
\label{firstpage}
\pagerange{\pageref{firstpage}--\pageref{lastpage}}
\maketitle

\begin{abstract}
  We use the Millennium simulation to show that halo clustering varies significantly with cosmic web type. Halos are classified as node, filament, sheet and void halos based on the eigenvalue decomposition of the velocity shear tensor. The velocity field is sampled by the peculiar velocities of a fixed number of neighbouring halos and spatial derivatives are computed using a kernel borrowed from smoothed particle hydrodynamics. The classification scheme is used to examine the clustering of halos as a function of web type for halos with masses larger than $10^{11}$. We find that node halos show positive bias, filament halos show negligible bias, and void and sheet halos are anti-biased independent of halo mass. Our findings suggest that the mass dependence of halo clustering is rooted in the composition of web types as a function of halo mass. The substantial fraction of node type halos for halo masses $\gtrsim 2\times10^{13}\,h^{-1}\rm M_\odot$ leads to positive bias. Filament type halos prevail at intermediate masses, $10^{12} - 10^{13}\,h^{-1}\rm M_\odot$, resulting in unbiased clustering. The large contribution of sheet type halos at low halo masses $\lesssim 10^{12}\,h^{-1}\rm M_\odot$ generates anti-biasing.
\end{abstract}

\begin{keywords}
cosmology: dark matter - large-scale structure of Universe -- methods: numerical
\end{keywords}


%
\section{Introduction}
Within the past few decades dark matter N-body simulations have contributed significantly to exploration of theories of cosmological structure formation \citep[e.g.,][]{Centrella1983, Klypin1983, Davis1985, Frenk1988, Springel2005b, Frenk2012}. Comparison of N-body halo catalogues with galaxy redshift surveys such as 2dF \citep{Colless2001} \& SDSS \citep{York2000} has led to a basic understanding of the connection between galaxy properties and underlying dark matter distribution. In particular, it is now established that galaxy clustering, as commonly measured by the two point correlation function, is determined by the clustering of the host halo population. The clustering of halos is thought to be largely independent of baryonic physics \citep{vanDaalen2014, Hellwing2016} and can therefore be studied by cold dark matter N-body simulations.

The clustering strength of halos is first and foremost a function of halo mass, i.e. high mass halos are more strongly clustered compared to low mass halos. This behaviour can be derived from the statistics of the initial density field \citep{Kaiser1984, Bardeen1986, Cole1989, Mo1996}. On the other hand, halo mass is correlated with shape, concentration, content, spin and formation time of the halos \citep{Frenk1988, Dubinski1991, Franx1991, Warren1992, Cole1996, Bullock2002, Jing2002, Bailin2005, Allgood2006, Bett2007, Maccio2008}. In a seminal study \cite{Gao2005} found that low mass halos ($\lesssim 10^{13} M_\odot$) that have formed earlier exhibit higher degrees of clustering than equal mass halos which have formed at a later time. Subsequent studies showed that several other ``secondary'' halo properties when used to sub-divide the distribution of halos in a given mass bin lead to systematic differences in the clustering behaviour \citep{Gao2005, Harker2006, Wechsler2006, Croton2007, Gao2007, Allgood2006, Bett2007, Hahn2007a, Hahn2007b, Dalal2008, Li2008, Faltenbacher2010a}. In the literature, the dependence of halo clustering on secondary halo properties is subsumed as \emph{assembly bias}.

The large scale structure of the Universe and its network-like structure was first noted by \citep{Shandarin1983}. It was later coined by \cite{Bond1995} as the cosmic web consisting of voids, sheets, filaments and nodes. TVoids are large scale underdensities which occupy the majority of the volume of the Universe. Their shapes are best described as convex irregular polyhedra. Voids are delineated by two dimensional sheets which show moderate overdensities. Filaments are the common line segments where neighbouring sheets meet. Nodes are located at the intersections of filaments and have the highest densities. The cosmic web is characterised by the large scale density, velocity and tidal fields, correlating the properties of dark matter halos residing within them \citep[e.g.,][]{Bond1995, Colberg1999, Kauffmann1999, Altay2006, Faltenbacher2008}. Various algorithms now exist to classify the large scale structure of the Universe into its constituent web types \citep[e.g.,][]{Aragon-Calvo2007a, Hahn2007a, Forero-Romero2008, Aragon-Calvo2010a, Falck2012, Hoffman2012, Cautun2013, Cautun2014}.

The degree of bias of halo clustering has been shown to depend on not only mass but also on other factors such as the anisotropy of the velocity field, formation time as well as shape and spin of halos \citep{Gao2005, Faltenbacher2010a, Gao2007}. In this analysis we introduce the \emph{cosmic web type of the halo} as sub-dividing criterion for an assembly bias study. The classification scheme used within is based on the velocity shear tensor \cite{Hoffman2012} which has been employed previously \citep[e.g.,][]{Libeskind2013} to investigate the correlations of halo properties with web-type.

Using publicly avaialble halo catalogues from Virgo - Millennium Database\footnote{http://gavo.mpa-garching.mpg.de/Millennium/}, we describe the dependence of halo clustering on web type. For that puropse each halo is assigned a cosmic web type based on the eigenvectors of the velocity shear tensor extracted from the velocities of the neighbouring halos \citep[cf.,][]{Hoffman2012, Fisher2016}. The outline of this paper is as follows: we review the characterisitics of the Millennium simulation and introduce the web type classification procedure in Sec.~\ref{sec::methods}. We then examine the clustering properties based on the halo-matter two point cross-correlation function for the different web types in Sec.~\ref{sec::cross_corr}. Sec.~\ref{sec::conclusion} presents some concluding remarks.
\section{Method}
\label{sec::methods}
In this section we give some details about the Millennium simulation used for this analysis. We decided to use the original Millennium simulation, with slightly outdated cosmological paramteres, for two reasons: 1) our results are generic, i.e. they do not depend on the exact cosmological model; 2) the comparison the seminal work by \cite{Gao2005} on assembly bias is straight forward. In the second part of this section, auto and cross-correlation functions are reviewed. Finally, core elements of the Lagrangian web classification algorithm introduced by \cite{Fisher2016} are discussed to the extent needed for the understanding of the analysis.
\subsection{Simulation \& Halo Sample}
The following analysis is based the Millennium simulation \citep{Springel2005b}. The first in a series of publicly available, high resolution, cosmological N-body simulations performed within the Millennium Simulation Project \citep{Lemson2006}. For the original Millennium Simulation Run a $\Lambda$CDM cosmological model with the following parameters has been adopted: $\Omega_{dm} = 0.205$; $\Omega_{b} = 0.045$; Hubble parameter $h_{100} = 0.73$; primordial power spectrum index $n=1$ and $\sigma_{8} = 0.9$. The evolution of $2160^3$ dark matter simulation particles in a 500 Mpc/h cube is followed resulting in a mass resolution of $8.6\times 10^8\, h^{-1}M_{\odot}$.

To identify the dark matter halos, a friends-of-friends (FOF) \citep{Davis1985} group finder with a linking length b=0.2 is employed and a minimum number limit of 20 particles is imposed. FOF groups are then decomposed into a distinct set of self-bound subhalos using \textsc{SUBFIND} \citep{Springel2001}, the most massive of which is then assigned as the main halo. The current analysis will be restricted to main halos.
\subsection{Quantifying Halo Clustering: The Two-Point Correlation Function}
The degree of clustering is quantified using the spatial \emph{two - point correlation} function \citep[e.g.,][]{Peebles1980}. Commonly, one distinguishes between two variations of the two point correltaion function: (i) the two-point \emph{auto}-correlation function which compares counts of pairs of points drawn from the same sample with the counts drawn from a corresponding random sample as a function of pair separation, while (ii) the two-point \emph{cross}-correlation function compares counts of pairs drawn from a main and a reference sample with the counts from the main and a corresponding random sample as a function of pair separation. In this paper, we utilise halo samples, defined by mass and web type, as main samples ($Q$-sample) and an a random subset of $\sim 10$ million simulation particles as reference sample ($R$-sample). Thus, the $R$-sample represents the overall matter distribution. Due to the simple geometry of the simulation box, pair counts involoving the random sample ($\mathcal{R}$-sample) can be determined analytically.

In the following we utilise the auto-correlation for the $R$-sample which reflects the clustering behaviour of the overall matter distribution and the matter-halo cross-correlation function. The relation between the two correlation functions is comonly used to determine the bias factor of the halo sample. The auto-correlation function of the $R$-sample is computed by:
\begin{equation}
\label{auto}
 \xi(r)_{mm} =  \frac{RR(r)}{\mathcal{RR}(r)} - 1\ ,
\end{equation}
where $RR (r)$ indicates the number of pairs from the reference sample with a spatial separation of $r$. $\mathcal{RR}(r)$ is the corresponding number of pair counts drawn from a random sample which is determined analytically. The halo-matter cross-correlation is computed for pairs from the $Q$ and the $R$ sample:
\begin{equation}
  \label{cross}
  \xi(r)_{hm} =  \frac{QR(r)}{Q \mathcal{R}(r)} - 1\ ,
\end{equation}
where $QR(r)$ are the mixed pair counts drawn from the halo sample ($Q$) and the reference sample ($R$). As for the auto correlation function, the correctly normalised counts for the main and the random sample, $Q \mathcal{R} (r)$, are computed analytically.

It is worth mentioning, that the cross-correlation functions, discussed here, can be interpreted as the \emph{averaged density contrast profiles}, scaled by the cosmic mean density, for a given halo sample:
\begin{equation}
  \label{contrast}
\delta= {\rho \over \rho_{\rm mean}} -1\ ,
\end{equation}
where $\rho$ is the local and $\rho_{\rm mean}$ cosmic mean density. This is a consequence of the fact that our $R$-sample reflects the overall matter distribution. Consequently, the halo-matter cross-pair counts for a given distance bin correspond to the cumulative mass found in all shells centered on the halos of that sample. The division by the halo-random cross-pair counts for the same shells is effectively a normalisation by the mean cosmic density. Finally, subtraction of one gives the density contrast.

We determine the cross-correlation functions on distance scales from 0.1 $h^{-1}$Mpc to 50 $h^{-1}$Mpc. The amplitudes at distances $\lesssim 2 h^{-1}$Mpc (depending on the mass of the halo) reflect the density within the halo. At distances of $\sim 10 h^{-1}$Mpc outside of the halo volumes, the comparison of the halo - matter cross-correlation function, $\xi(r)_{hm}$ (Eq.~\ref{cross}), and the matter auto-correlation function, $\xi(r)_{mm}$ (Eq.~\ref{auto}), determines the clustering strength of the halo sample relative to the overall clustering of matter. If the amplitudes of the halo correlation function are larger than that of the mass correlation function, the halo sample is said to show positively biased clustering. The reverse behaviour is referred to as anti-biased halo clustering.

\subsection{Cosmic Web Classification Algorithm}
Classifying individual halos into web types is done using the Lagrangian classifier given by \cite{Fisher2016} applied to a given halo sample. This is accomplished by constructing the velocity shear tensor \citep[see][]{Hoffman2012} using techniques borrowed from smoothed particle hydrodynamics (SPH) given as:
\begin{equation}
  \frac{\partial \mathbf{v}^{\alpha}_{s,i}}{\partial {\beta}} \simeq \frac{1}{n_i}\sum_{j=1} ^N (\mathbf{v}^{\alpha}_i - \mathbf{v}^{\alpha}_j)\frac{\partial}{\partial \beta} W(|\mathbf{r}_i-\mathbf{r}_j|, h_i)\ .
  \label{grad_component}
\end{equation}
where the Greek indices refer to the coordinate system $(x, y, z)$ and the Latin indices represent the halo labels, $n_i$ is the halo number density at the location of halo $i$\footnote{Note: More commonly, SPH approaches utilise the density, $\rho_i$, instead of the number-density, $n_i$, used here. This requires to include the masses of the particles (halos), $m_j$, within the summation. Since we interpret halos as test particles of the underlying velocity field we bypass the reliance on generally ill-determined halo masses by using $n_i$ instead of $\rho_i$.}, $\mathbf{v}$ are the halo velocities and $W(|\mathbf{r}_i-\mathbf{r}_j|, h_i)$ is the smoothing kernel. We adopt a cubic spline kernel \citep{Monaghan1985}, given by \cite{Springel2010} as:
\begin{equation}
  w(q)= \frac{8}{\pi h^3}
  \begin{cases}
    1-6q^2+6q^3,\qquad  & 0 \le q \le \frac{1}{2} \\
    2(1-q)^3, \qquad  & \frac{1}{2} < q \le 1\\
    0, \qquad & \text{ Otherwise }
  \end{cases}
  \label{kernel}
\end{equation}
with $q=r/h$ and $W(r,h)=w(q)/h^3$. The adaptive smoothing length, $h$, is set individually to ensure exactly 32 neighbours are accounted for. The adaptive approach largely prevents local density and velocity field sampling errors.

Once the 9 elements of the velocity shear tensor have been constructed according to Eq.~\ref{grad_component}, the velocity shear tensor is diagonalised to obtain its eigenvalues which are then sorted. Defining an eigenvalue threshold value $\lambda^{th}$ allows one to classify the halo web type by counting how many of its eigenvalues are greater than the threshold value. Nodes, filaments, sheets and voids have 3, 2, 1 and 0 eigenvalues greater than the threshold value respectively.

An eigenvalue greater than the threshold value is interpreted as a collapse along its corresponding eigenvector \citep{Zeldovich1970, Hoffman2012}. Expansion along a given eigenvector occurs when the corresponding eigenvalue is less than the threshold value. Nodes collapse and voids expand along all 3 axes. Sheets collapse along one of its axis and expand along two axes while filaments are observed to collapse along two axes and expand along the third. In this work we adopt a threshold value of zero.

\cite{Hoffman2012} considered the threshold value as a free parameter and determined visually as $\lambda^{th} = 0.44$. \cite{Fisher2016}, based on a Lagrangian approach for the determination of the velocity shear tensor, showed that adding the Hubble flow to the peculiar velocity field (Hoffman's analysis is based on the peculiar velocity field) allows one to set $\lambda^{th} = 0$. Zero seems to be the most natural discriminator between contraction ($\lambda^{th} > 0$) and expansion ($\lambda^{th} < 0$) along the eigenvectors of the velocity shear tensor.

\subsection{Hubble Flow}
One subtle aspect of the classification algorithm that needs clarification is the in/exclusion of the Hubble flow when sampling the velocity field for the subsequent determination if the velocity shear tensor. It has been shown in \cite{Fisher2016} that the exclusion of the Hubble flow contaminates the classification scheme by the artificial shrinking of virialised structures. Therefore, inclusion of the Hubble flow is particularly important if the classification scheme is applied to dynamical information from \emph{within virialised structures}. This is the case if simulation particles or subhalo populations (i.e. not only main halos) are employed for sampling the velocity field.

On the other side, if one focuses on the classification of the main halos based on the large scale environment outside of virialised structures the above mentioned shrinking of the virialised regions due to the exclusion of the Hubble flow does not contaminate the classification scheme. On the contrary, the restriction of the sample to main halos causes sph-kernels to extend to rather large distances where the Hubble expansion begins to dominate. In this case the inclusion of the Hubble flow causes a problem as for distances $\gtrsim 5$ Mpc the ambient main halos show an outward drift due to the Hubble flow resulting in a bias of the classification scheme towards more 'expanding environments', for instance a filament my be classified as sheet or void or a sheet may be determined as void. The expansion measured this way is not a result of the local gravitational field.

The classification scheme used for the present analysis employs the peculiar velocities of main halos \emph{without including the Hubble flow} to reveal the impact of gravity on the large scale velocity field. The utilisation of peculiar velocities in combination with the sampling of the velocity field over relatively large distances allows the threshold value for the classification to be set to $\lambda^{th} = 0$. This approach produces number fractions for the halo - web type correlations comparable to \cite{Hahn2007a}. A zero threshold value permits a natural interpretation as positive and negative eigenvalues correspond to gravitational contraction and expansion respectively.

\section{Results \& Discussion}
\label{sec::cross_corr}
The results reported here are based on the main halo sample with top-hat masses above $ 9 \times 10^{10}\, h^{-1} {\rm M}_{\odot}$, where top-hat mass refers to the mass within the radius where the halo has an overdensity corresponding to the value at virialisation in the top-hat collapse model for the adopted cosmology. The peculiar velocities of the these halos are used to sample the large scale velocity field and to determine the velocity shear tensor at the halo locations. According to the number of positive and negative eigenvalues, the halos are classified as node, filament, sheet and void halos. In the first part of this section we discuss the overall number statistics of the various halo types. The second part compares the matter cross-correlation functions of halo samples for the different cosmic web types. Finally, we present the mass dependence of the clustering of halos for a given web type. Therefore, the clustering behaviour is investigated for halos in narrow mass bins \citep[same as used in ][]{Gao2005}.

\subsection{Classification results}
According to the classification scheme discussed above, the web type number fractions for the classification of the Millennium main halo sample (at $z = 0$), including all main halos with top-hat masses $\geq 9 \times 10^{10}\, h^{-1} {\rm M}_\odot $, are: 9.15\% node, 62.61\% filament, 27.55\% sheet and 0.91\% void halos. For the simulation output at $z = 1$ using all main halos with top-hat masses $\geq 5 \times 10^{9}\, h^{-1} {\rm M}_\odot$ (the scaling factor between the two mass limits (at $z=0$ and $1$) is obtained by comparing the characteristic masses of the spherical collapse model at the two redshifts, also see Figure in the Appendix) the corresponding number fraction are: 10.22\% node, 64.76\% filament, 24.51\% sheet and 0.91\% void halos. Filament halos constitute the largest subset of halos in the mass range considered here. The statistics indicate a skewness towards web types with a larger number of positive eigenvalues of the velocity shear tensor. It is suggestive that environments of higher infall dimensionality (ID), such as nodes (ID=3) and filaments (ID=2), are more conducive to the formation of halos. On the other hand, one does not expect to find many halos in void region as shear forces are causing an expansion along all 3 dimensions (ID=0). The relatively small number of node halos indicates that only a small volume fraction of the universe shows infall along all three dimensions.

\begin{figure}
 \centering
 \includegraphics[width = 0.45\textwidth]{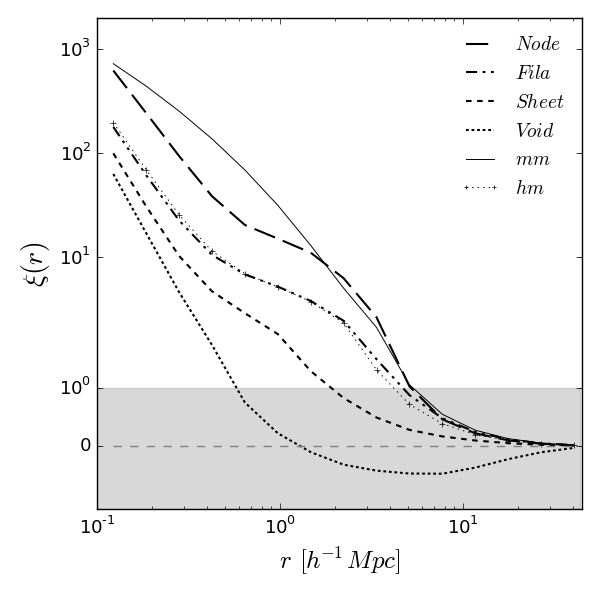}
 \caption{The halo-matter cross-correlation functions per web type compared to the matter auto-correlation function (solid line) and the halo-matter cross-correlation for the entire halo sample (thin dotted line with '+' markers) at present time, $z=0$. The velocities of all halos with spherical collapse top-hat masses $\geq 9 \times 10^{10}\, h^{-1} {\rm M}_\odot $ are utilised for the web-classification scheme. The matter distribution is sampled by a subset of ~10 Million randomly selected simulation particles. Note: the y-scale is logarithmic for $\xi > 1$ and linear for $\xi < 1$ indicated by the shaded region. The dashed gray line indicates $\xi = 0$. A value of $\xi < 0$ indicates less clustering than the random sample. Nodes (large dashes), filaments (dash-dot), sheets (small dashes) and voids (dotted) exhibit very different clustering behaviour. The effect is largely tiered, with the web types of higher infall dimensionality demonstrating significantly more clustering. Voids are under-clustered at larger scales.}
 \label{2pcf_all}
\end{figure}
\subsection{Halo clustering dependence on web type}
Fig.~\ref{2pcf_all} shows the two point cross and auto-correlation functions for the halo and the overall matter distributions, respectively. The matter distribution is sampled by a subset of ~10 Million randomly selected simulation particles (or ~0.1\% of the total sample). The classification is based on the velocity field sampled by all main halos with masses $\geq 9 \times 10^{10}\, h^{-1} {\rm M}_\odot $. We use a linear scale below $\xi = 1$ to be able to display cases where
under-clustering is occurring, i.e. where $\xi < 0$. The level of clustering varies significantly with web type as the cross-correlation functions are staggered according to the ID of the web type.

For small $r$ the halo-matter cross correlation functions effectively give the averaged density profiles of the halo sample ($Q$-sample). Thus, the tiered behaviour observed on small scales $\lesssim 1\,h^{-1}\rm Mpc$ reveals a size and mass segregation of the halos corresponding to web type. Void halos are preferentially found among the smallest, least massive halos in the entire sample. The opposite holds for node halos which occupy the largest and most massive halo ranges. Filament and sheet halos occupy the intermediate mass and size ranges.

The staggered arrangement observed at distances $\gtrsim 1\,h^{-1}\rm Mpc$ is a seamless continuation of the behaviour on small scales. At about 10$h^{-1}$ Mpc, where commonly the bias for a given halo population is measured, the comparison of the halo-matter cross-correlation functions with the matter auto-correlation function reveal that halo populations associated with lower ID web types exhibit larger anti-bias. Void halos are shown to be the least clustered web type. Interpreting the halo-matter cross-correlation function for the void halos as averaged density profiles suggests that they reside in under-dense environments of 10s of Mpc in size.

The thin dotted line with the '+' markers in Fig.~\ref{2pcf_all} shows the halo-matter cross-correlation function for the entire halo sample. One finds a noticeable similarity to the filament halo cross-correlation function which can be explained by the fact that filament type halos constitute the dominant fraction in the entire halo sample (62.61\%). The entire halo sample shows a marginal anti-bias which is a result of the mass distribution within the entire sample, i.e. in numbers the anti-biased halos at the lower end of the mass range dominate over the strongly biased halos at the high mass end.

The tiered behaviour of the halo-matter cross-correlation functions on all scales indicates that halos classified as web types with the highest IDs (IDs = 3 and 2 for nodes and filaments respectively) are usually found in higher density environments. Similarly, low ID (0 and 1 for voids and sheets respectively) corresponds to low density regions. Thus, as expected, there is a direct correlation between ID and density of the environment. Since our web type classification scheme is based on halo positions and velocities alone, this agreement is an independent confirmation of the good correspondence between the cosmic web classification based on density and that based on velocity shear reported by \cite{Cautun2014}.

So far the sampling of the velocity field, and ultimately the computation of the eigenvalues of the velocity shear tensor, is based on the entire halo sample, comprised of all main halos $\geq 9 \times 10^{10}\, h^{-1} {\rm M}_\odot$ in the simulation box. The determination of the velocity gradient uses the 32 nearest neighbour main halos. The most massive halos in the simulation show masses above $10^{15}\,h^{-1} M_\odot$. In general, it is expected that high mass halos are surrounded by a plethora of low mass halos (10 - 10000 smaller in mass). Thus, the velocity shear tensor associated with high mass halos may frequently be based on a set of low mass halos close to, but outside of, the virial radius.

These regions are know to be contaminated by  backsplash halos, i.e. halos which have been part of the main halo before being ejected \citep[e.g.,][]{Moore2004,Warnick2008}. To check the effect of the backsplash halos we have computed the shear tensor excluding neighbour halos within two times the virial radius and obtained very similar results, with the encouraging exception that the very high mass void halos disappear. In the following section, which presents the main results, we only use narrow mass bins for the sampling of the velocity field and calculation of the velocity shear tensor which suppresses the contribution of backsplash halos as they are typically orders of magnitude less massive than the main halo. Thus, we do not investigate the impact of backsplash halos any further.

\begin{figure}
 \centering
 \includegraphics[width = 0.45\textwidth]{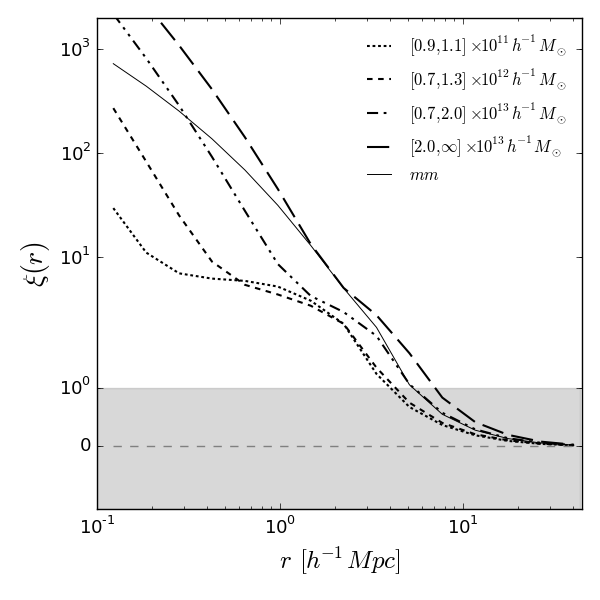}
 \caption{The cross-correlation functions for halos in each mass bin and the matter auto-correlation function at $z=0$. Higher mass halos are more strongly correlated relative to low mass halos and halos above the characteristic mass $M_*$ tend to be more bias while less massive halos exhibit more anti-bias.}
 \label{massonly}
\end{figure}
\begin{figure*}
 \centering
 \includegraphics[width = 0.85\textwidth]{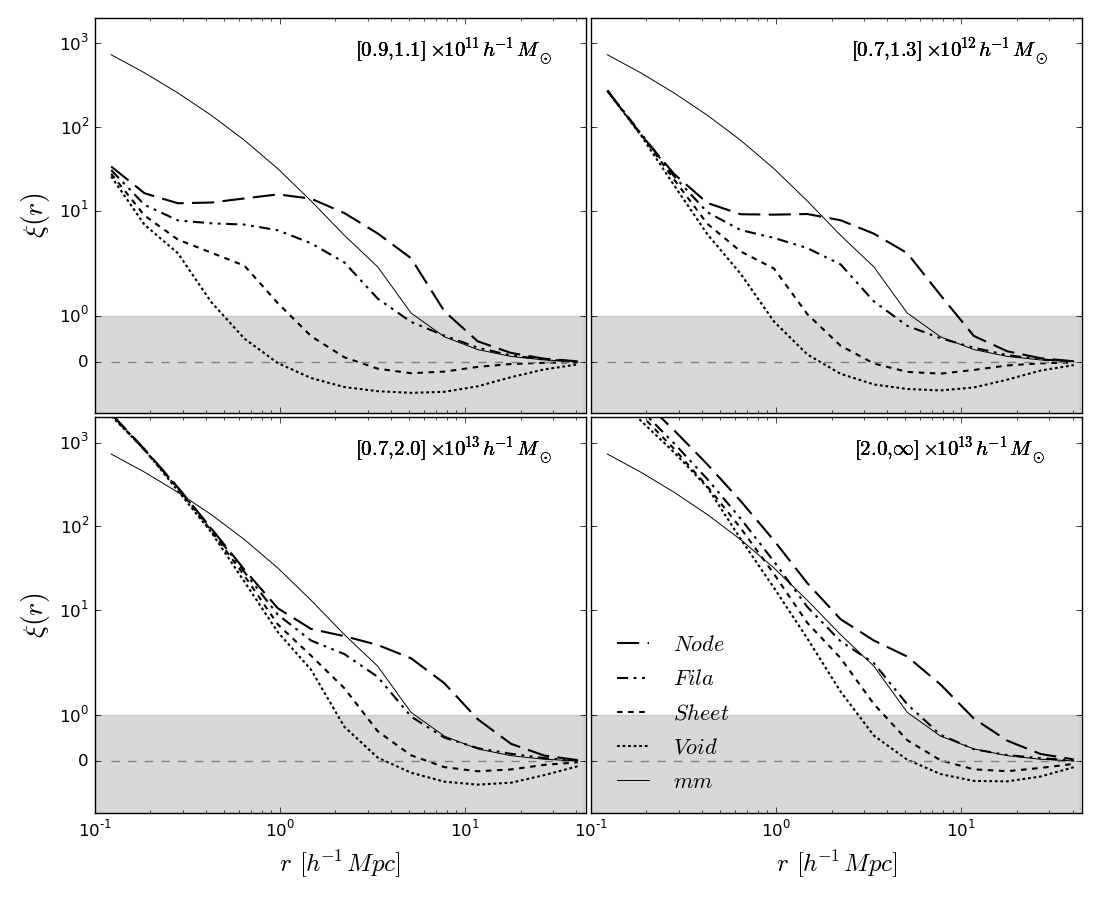}
 \caption{The halo-matter cross-correlation functions per web type and the matter auto-correlation function in four increasing mass bins for $z=0$. Clustering is seen to vary significantly with web type for all mass bins. Low mass halos differ considerably in size, density and exhibit a large anti-bias on smaller scales. For all masses at large radii, void and sheet halos are under-clustered and node halos are consistently bias. The $z=1$ results are given in the Appendix, Fig.~\ref{2pcf_mbz1}.}
 \label{2pcf_mb}
\end{figure*}
\subsection{Mass Dependence of Web Type Clustering}
\label{sec:main}
Akin to the work in \cite{Gao2005}, we examine the mass dependence of halo clustering in four narrow mass bins. Within each mass bin we subdivide the halo sample according to web type and investigate the differences in the associated halo-matter cross-correlation functions. The sampling of the velocity field and consequently the determination of the cosmic web type is confined to halos in the same mass bin. For the highest mass bin the velocity shear tensor is computed based on halos more massive than $ 2 \times 10^{13}\,h^{-1} \rm M_\odot$ resulting in an average distance to the 32nd neighbour of $35\,h^{-1}\rm Mpc$. The confinement to narrow mass bins takes the different sizes of the halos into account, i.e. the determination of the web type of low mass halos is based on low mass neighbour halos which on average have smaller separations, whereas for the determination of the web type of high mass halos only the halos in the high mass bin are used which show on average larger separations. Somewhat counterintuitive is the appearance of a small fraction (0.3\%) of void halos in the highest mass bin. This average expansion of the ambient halo-velocity filed as reliably measured by the SPH-kernel may be caused by halos of similar size which just experienced a close encounter and presently are on a separation course or by the gravitational impact of massive structures nearby ($\sim 35\,h^{-1}\rm Mpc$).

As a cross check, we have used the entire halo sample to compute the web types for the halos in the various mass bins and find qualitative agreement despite the contamination by backsplash halos in this case. Since the former approach determines web type and clustering based on the same halo sample we have decided to present the results of this, methodologically more consistent approach.

Fig.~\ref{massonly} shows the halo-mass cross-correlation functions for all halos in the indicated mass bins, again on linear scales for $\xi(r)\le1$. As expected, clustering increases with mass. The correlation functions are tiered. The charatersitic mass derived from the sperical collapse model within the adopted cosmology is $M_*(z=0) approx 7 \times 10^{12}h^{-1} M_{\odot}$. In the lower mass bins, less than the characteristic mass, a anti-bias, measured at scales from $5$ to $20\,h^{-1}\rm Mpc$, can be observed. Conversely, the highest mass bin is defined greater than $M_*$ and one observes a large bias factor. Halos in the second highest bin have masses most similar to $M_*$ and they display very little bias.

In Fig.~\ref{2pcf_mb}, we calculate the web type clustering per mass bin. All panels of Fig.~\ref{2pcf_mb} display the same behaviour as in Fig.~\ref{2pcf_all} whereby the web type cross-correlation functions are tiered by web type. At small scales, the halo-matter cross-correlation function measures the average density profiles of the halo sample and it can be seen that the profiles converge independent of web type. The deviation seen in the highest mass bin is a result of the large mass range covered by this mass bin, which causes a behaviour reminiscent of Fig.~\ref{2pcf_all} where the correlation function for the entire mass range is displayed.

Arguably the most striking feature in Fig.~\ref{2pcf_mb} is the similarity of the correlation functions for a given web type independent of mass. In all four mass bins the node halos are strongly biased at scales $\gtrsim 5\,h^{-1}\rm Mpc$, the filament halos follow the matter auto-correlation function closely and sheet and void halos are strongly anti-biased with the zero-crossing occurring at much smaller scales, when compared to the entire halo sample. The exact scale of the zero-crossing seems to be influenced by the average halo radius in the specific sample but the maximal negative amplitudes are very similar for a given web type independent of mass.

On linear scales, $\sim 10 h^{-1}$ Mpc, the clustering of low mass halos ($ \lesssim 10^{13}M_{\odot} $) has been shown to depend substantially on halo formation time \citep{Gao2005}. The clustering of halos which currently have the same mass depends on their formation time. Halos that assembled at high redshift are substantially more clustered than those that assembled more recently. Subsequently, many other properties of halos, such as concentration and velocity anisotropy parameter, have been shown to have an impact on halo clustering besides the halo mass (see introduction). Fig.~\ref{2pcf_mb} clearly demonstrates that the web type of halos also has a strong impact on halos clustering. Very simular conclusions can be drawn from the correponding $z=1$ results displayed in Fig.~\ref{2pcf_mbz1} in the Appendix.

\begin{figure}
 \centering
 \includegraphics[width = 0.5\textwidth]{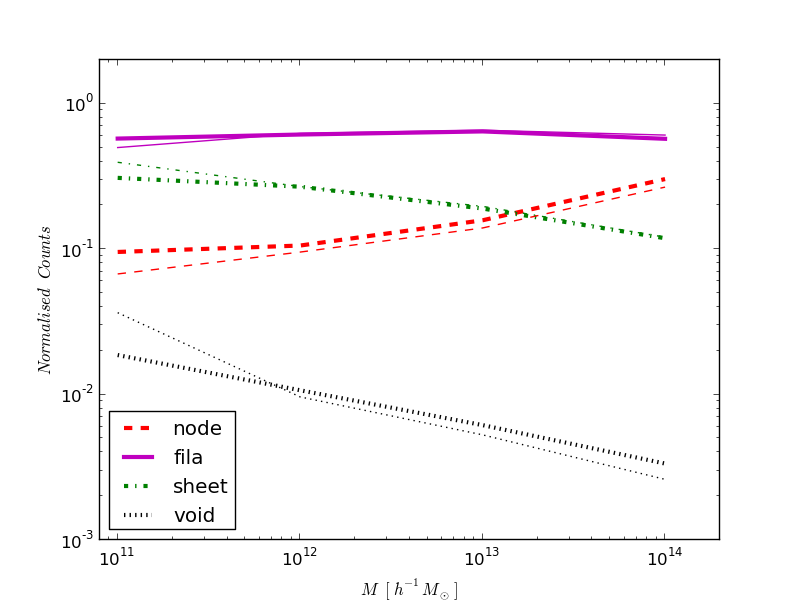}
 \caption{Normalised halo web type fractions as a function of halo mass. The highest mass bin has an infinite upper limit so the abscissa values for these points are chosen arbitrarily. Filament halos dominate for all halo masses with a relatively high contribution of sheet halos at lower masses. For the highest mass bin, node halos account for a substantial fraction of the overall halo population. Void halos do not contribute more than 1\% at any mass bin. The bold lines give the results for $z = 0$ and the thin lines display the fractions for $z = 1$, where the mass bins have been rescaled accounting for the reduction of the characteristic mass $M_*(z=1)\approx 3.5 \times 10^{11} h^{-1}M_\odot$.}
 \label{webmassfrac}
\end{figure}
\subsection{Dependence of Web Type Fractions on Mass}
The above analysis has revealed that same mass halos show different clustering behaviour depending on their web type. In order to gain insight into the mass dependence of the halo-matter cross-correlation functions for the entire halo sample, as seen in Fig.~\ref{massonly}, we now investigate which web types are responsible for the overall clustering per mass bin.

Fig.~\ref{webmassfrac} illustrates the web type fractions per mass bin (for visual impression, the last bin mass value is set to $10^{14}$, otherwise plotted using the midpoint of the respective bin). The fraction of halo web types is largely mass dependant. Node halo fractions are monotonically increasing with mass so that in the highest mass bin they account for the second largest fraction of all halos, followed by sheets and voids. Filaments are dominant in all mass bins. In the low mass regime, filament halos are seen to occupy the largest fraction of halos followed by sheet halos. Sheet and void fractions decrease with mass, the latter being less than 1\% for all mass bins.

Fig.\ref{webmassfrac} suggests that filaments and sheets determine the overall clustering in the lower mass ranges ( $\sim 10^{11}$ to $\sim 10^{12}\, h^{-1}\rm M_\odot$)  causing anti-bias. For intermediate mass ranges ( $\sim 10^{12}$ to $\sim 10^{13}\, h^{-1}\rm M_\odot$), filament halos dominate which results in marginal bias. Finally, the substantial fraction of node halos along with filament halos in the high mass ranges ($\gtrsim 5 \times 10^{14}\, h^{-1}\rm M_\odot$) generates positively biased halo clustering. The thin lines in Fig.\ref{webmassfrac} display the web type fractions for $z = 1$, where the mass bins are rescaled by the ratio of the characteristic masses of the spherical collapse model, $M_*(z=0)\approx 7 \times 10^{12} h^{-1}M_\odot$ and $M_*(z=1)\approx 3.5 \times 10^{11} h^{-1}M_\odot$. The agreement between $z=0$ and $z=1$ shows that the contribution of web types to the overall halo clustering is relatively constant over time.

\section{Conclusions}
\label{sec::conclusion}
We have employed main halos, with top-hat masses $\geq 9 \times 10^{10}\,h^{-1}\rm M_\odot$, to sample the cosmic velocity field and compute the velocity shear tensor. Based on the infall dimensionality (i.e. the number of positive eigenvalues , indicative of infall, for the associated velocity shear tensor) halos are classified into cosmic web types, namely node, filament, sheet and void halos. With the classification scheme in place, the clustering of different web type halos can be analysed.\\

Our main findings are:\\\\
1) The overall number fractions are: 9.15\% node, 62.61\% filament, 27.55\% sheet and 0.91\% void halos. The expansion in all three directions inhibits the formation of void halos. Node halos are rare because a converging velocity field may be restricted to very small volumes. Filament and sheets occupy a relatively large cosmic volume and can provide sufficient matter supply to host a large number of halos.\\\\
2) The halo-matter cross-correlation functions for different web types show tiered behaviour on all distance scales probed (0.1 - 50 $h^{-1}$ Mpc). There is a direct correlation between infall dimensionality and density of the environment, i.e. on average nodes, filaments, sheet and void halos live in environments with decreasing density. The velocity shear tensor based on halo velocities allows for a credible determination of halo web types.\\\\
3) The halo-matter cross-correlation functions for different web types computed in narrow mass bins: $~\sim 10^{11}$; $~\sim 10^{12}$; $~\sim 10^{13}$ and $> 2 \times 10^{11}\,h^{-1}\rm M_\odot$ reveal strong differences in the bias of the considered halo sub-sample. Node halos in all mass bins show positive bias. Filament halos in all mass bins display marginal bias on scales from 5 to 50 $h^{-1}$ Mpc. Sheet and void halos are anti-biased in all mass bins. The clustering behaviour of the different web types seems almost independent of mass.\\\\
4) Motivated by the findings reported under point 3, we determined the number fraction of halos of a given web type per mass bin in order to understand the structure of the clustering for the entire halo sample. We find that filament and sheet halos dominate at the low mass range ($10^{11} - 10^{12}\,h^{-1}\rm M_\odot$) causing a marginal anti-bias of the entire halo sample. In the mass range $10^{12} - 10^{13}\,h^{-1}\rm M_\odot$ the contribution of sheet halos decreases and node halos become more frequent. This is reflected in the diminishing bias of the halo clustering in this mass range. Finally, for halo mass above few times  $10^{13}\,h^{-1}\rm M_\odot$ node and filament halos prevail causing positive bias of the overall halo sample.\\

In conclusion, the clustering analysis for halos subdivided by mass and web type provides general insight into halo clustering. The relative contribution of halos of a given web type is dependant on halo mass and has a strong impact on the overall clustering behaviour.

\section*{Acknowledgements}
We are grateful for the comments of the anonymous referee which helped to improve the quality of the paper. The authors acknowledge the use of data from the publicly available Millennium Simulation (\url{http://www.mpa-garching.mpg.de/millennium}), carried out by the Virgo Consortium. We would also like to acknowledge the support from the National Astrophysics and Space Science programme (NASSP).




\bibliographystyle{mnras}

%


\appendix
\begin{figure*}
 \centering
 \includegraphics[width = 0.85\textwidth]{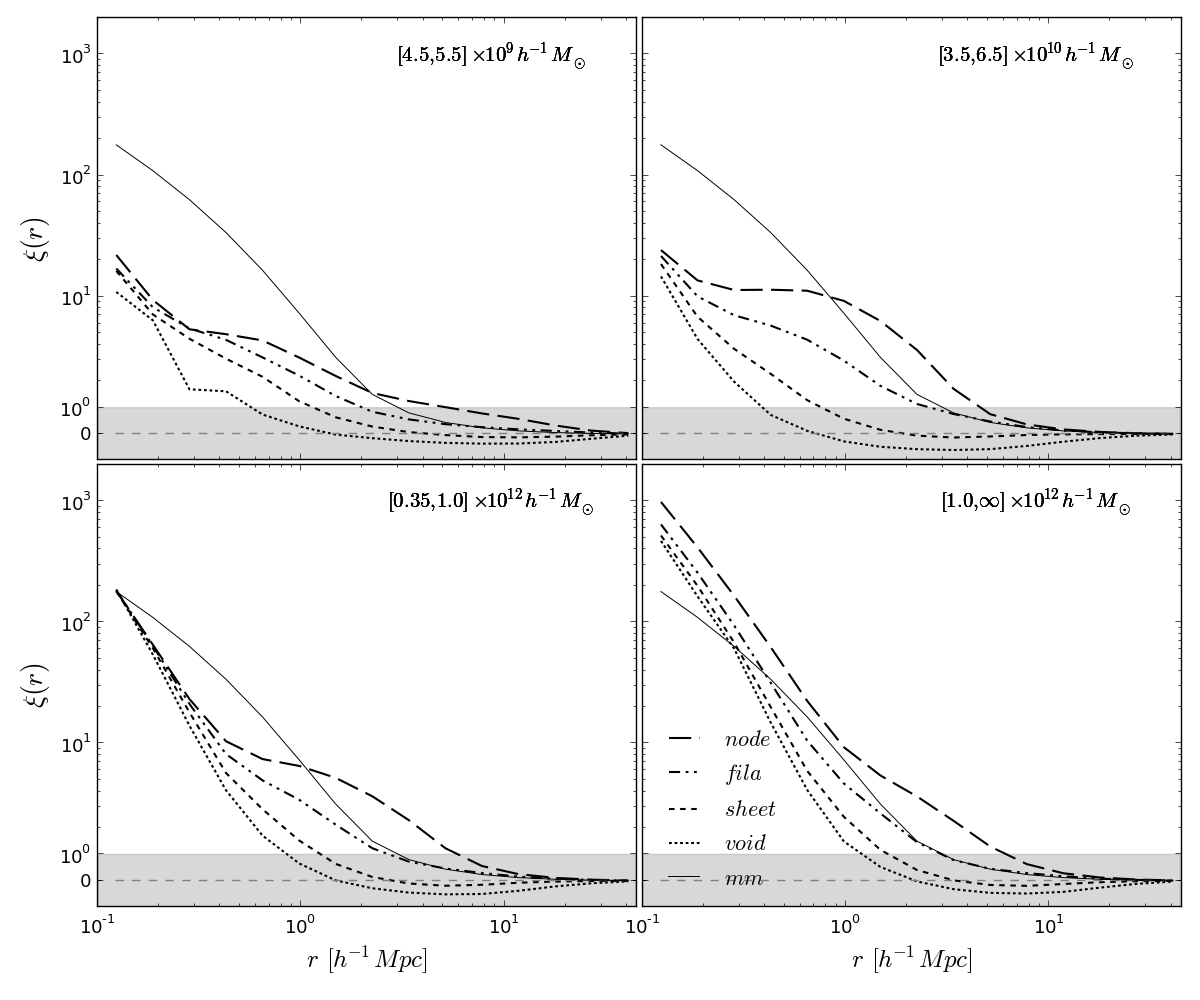}
 \caption{The halo-matter cross-correlation functions per web type and the matter auto-correlation function in four increasing mass bins. Same plots as shown in Fig.~\ref{2pcf_mb} but for a snapshot of redfshift $z = 1$. Compared to the $z=0$ results the mass bins have been rescaled by the ratio of the characteristic masses of the spherical collapse model, $M_*(z=0)\approx 7 \times 10^{12} h^{-1}M_\odot$ and $M_*(z=1)\approx 3.5 \times 10^{11} h^{-1}M_\odot$. Apart from an overall downscaling of the correlation functions due to the earlier evolutionary stage, all structural features of the halo-matter cross-correlation functions compare very well for $z=0$ and $z=1$ results.}
 \label{2pcf_mbz1}
\end{figure*}

%
%
%
%

\bsp	
\label{lastpage}
\end{document}